\documentclass[oribibl]{llncs}

\usepackage[english]{babel}
\usepackage[utf8x]{inputenc}
\usepackage{amsmath}
\usepackage{graphicx}
\usepackage{url}
\usepackage{multirow}
\usepackage{epstopdf}
\usepackage[colorinlistoftodos]{todonotes}

\begin{document}

\mainmatter
\title{Large scale citation matching using Apache~Hadoop}

\author{Mateusz Fedoryszak \and Dominika Tkaczyk \and Łukasz Bolikowski
}
\authorrunning{Mateusz Fedoryszak et al.} 
\institute{Interdisciplinary Centre for Mathematical and Computational Modelling,\\
University of Warsaw\\
\email{\{m.fedoryszak, d.tkaczyk, l.bolikowski\}@icm.edu.pl}}

\maketitle  

\begin{abstract}
During the process of citation matching links from bibliography entries to referenced publications are created. Such links are indicators of topical similarity between linked texts, are used in assessing the impact of the referenced document and improve navigation in the user interfaces of digital libraries. In this paper we present a citation matching method and show how to scale it up to handle great amounts of data using appropriate indexing and a MapReduce paradigm in the Hadoop environment.
\end{abstract}

\section{Introduction}
\label{sec:introduction}

Since Hitchcock et al. \cite{HitchcockCHHH:1997} demonstrated a proof-of-concept system that performed
autonomous linking within Cognitive Science Open Journal, the problem of citation matching (i.e. linking citation strings referencing the same paper) has been tackled in countless papers by means of various methods \cite{Fedoryszak2013}.


Huge interest in a citation resolution is not a surprise as it is a fundamental step in creating a digital library of scholarly publications. Having relationships between documents conveying the fact that document A references document B allows to provide more user-friendly interfaces \cite{HitchcockCHHH:1997}, perform scientometrical analysis \cite{Garfield:2006,Hirsch2005} and link-based classification \cite{Bolelli2006,MacskassyP:2007}.

Considering the rapid growth of the number of scientific publications, we need to seek new ways of dealing with large amounts of data. Recently, a MapReduce paradigm \cite{DeanG:2004} and Apache Hadoop, its open-source implementation, have been gaining popularity. It has already been used for entity matching by Paradies et al. \cite{Paradies2012}.

In this paper we present our own approach to citation matching in Hadoop environment. We start by demonstrating a small scale matching method in Section \ref{sec:small_scale_disambiguation}. An interesting author similarity measure is presented in Section \ref{sec:author_matching}. A~very important part in the scalability of our solution is played by an approximate index for heuristic matching, which is described in Section \ref{sec:heuristic_author_indexing}. Finally, in Section~\ref{sec:hadoopisation}, some details of index building and actual citation matching implementation utilising MapReduce paradigm are revealed.

\section{Small scale disambiguation}
\label{sec:small_scale_disambiguation}

First of all, let us look on how reference disambiguation is performed when working on a small scale. As we do not need to deal with great amounts of data here, we can afford to count the similarity between every citation string pair. Having pairwise similarities, we can apply any clustering algorithm. In our tests, we have used a basic single-link algorithm.

Let us describe first things first, though. To begin with, we need to extract metadata from a given reference string. We will describe that in section \ref{sec:citation_parsing}. However, metadata extracted from references may be inaccurate or malformed. We have therefore developed measures of fussy match. They are described in section \ref{sec:metadata_fields_matching}. From match factor of particular metadata fields, we need to draw conclusions about the whole citation. SVM \cite{Cortes1995} is employed for that task. Finally, the accuracy evaluation is presented in section \ref{sec:evaluation}.

\subsection{Citation parsing}
\label{sec:citation_parsing}

One of citation matcher requirements is the access to the metadata of input citations.
Unfortunately, in some cases the matcher has to deal with citations in the form of raw strings. In
such situations citations need to be preprocessed in order to extract the required metadata. This 
is done by a citation parser, whose role is to identify fragments of the input citation string 
containing meaningful pieces of metadata information. The information we extract at this stage 
includes: an author, a title, a journal name, pages and a year of publication.

The parsing is performed in several steps. First, a reference string is tokenised into a list of
tokens, each of which is in one of the following forms: a string containing only letters, a string 
containing only digits, a string containing letters and digits, a single other character. After that 
the parser computes a set of feature values describing each token. The tokens represented by vectors 
of features are then classified into several categories that correspond to metadata fields. The token
classifier is the heart of the citation parser. The classifier is based on Conditional Random Fields
and is built on top of GRMM and MALLET packages~\cite{McCallum2002}.

We use 42 features to describe a token:
\begin{itemize}
\item features based on the presence of a particular character class, eg. digits, lowercase/uppercase
letters, Roman digits,
\item features checking if the token is a particular character (eg. a dot, square bracket, a comma or a 
dash),
\item features checking if the token is a particular word,
\item features checking whether the token is contained by the dictionary built from the dataset, eg. a
dictionary of cities or words commonly appearing in the journal title.
\end{itemize}

It is worth noticing that the class of a citation token depends in fact not only on its feature vector, 
but also on surrounding tokens. To make the classifier aware of this dependency, each feature vector 
is extended by adding the features of two preceding and two following tokens.

Citation parser is a part of CERMINE --- a metadata and content extraction tool \cite{TkaczykBCR:2012}.

\begin{figure}
\begin{center}
\includegraphics[width=\linewidth]{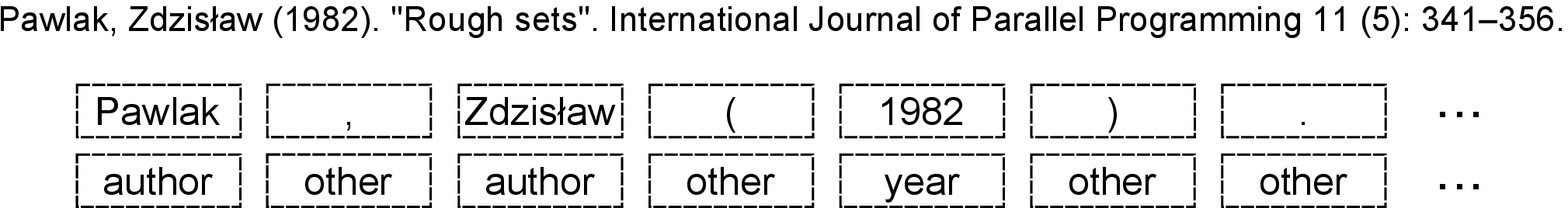}
 \end{center} 
\caption{A citation string, its tokens and token classes.}
\label{fig:parsed_citation}
\end{figure}

\subsection{Metadata fields matching}
\label{sec:metadata_fields_matching}

Not only do citations contain spelling errors, but also they differ in style which leads e.g. to differences in journal names abbreviating conventions. That is why even matching of parsed citations is not a trivial task. To address that we introduce the measures of similarity fitted to the specifics of various metadata fields. The overall similarity of two citation strings is obtained by applying a linear SVM using field similarities as features.

\subsubsection{Definitions}
\label{sec:definitions}

Let $trigrams(s)$ be a multiset of trigrams \cite{Ukkonen1992} from a string $s$. We define a \emph{trigram similarity} between strings $s$ and $t$ as $$sim_{tri}(s, t)=2\frac{|trigrams(s) \cap trigrams(t)|}{|trigrams(s) \cup trigrams(t)|}$$

\emph{Token similarity} is defined in a similar manner. Let $tokens(s)$ be a multiset of tokens from a string $s$. Then $$sim_{token}(s, t) = 2\frac{|tokens(s) \cap tokens(t)|}{|tokens(s) \cup tokens(t)|}$$

\subsubsection{Author matching}
\label{sec:author_matching}

Author names in citation string can take various forms. In some cases given names are abbreviated, sometimes they are placed before, sometimes after the surname. Additional titles may also be added. We have developed two ways of measuring author field similarity. The first, most basic one, computes token and trigram similarity. 

The second way of defining similarity is much more sophisticated. It is based on finding the heaviest matching of tokens, what can be seen as an instance of assignment problem \cite{Kuhn1955}. Each token can match at most one token in the other citation. Each pair of matched tokens has a weight assigned. The weight of a whole matching is the sum of weights of matched tokens. The weight of token pair is determined by a token similarity and their relative distance in the whole string. The distance component allows us to take into account the ordering of authors. It also lets us distinguish between papers authored by John Smith and Jane Doe and those written by Jane Smith and John Doe. The weights are assigned in a way that guaranties that the weight of the matching lays in [0,1] so it can be treated as similarity. The token similarity is measured in terms of edit distance, with just one exception: a pair of tokens, one of length lesser or equal 2 being a prefix of another, is assumed to have edit distance \cite{ElmagarmidIV:2007} equal 1.

\begin{figure}
\begin{center}
\includegraphics[scale=0.7]{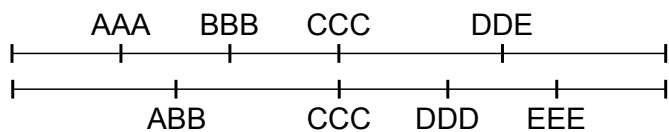}
 \end{center} 
\caption{\textbf{Token distances}. The length of citation is normalised to 1. Perfectly matching tokens are aligned to have distance equal 0. All other tokens are positioned relatively to them.}
\label{fig:distance_example}
\end{figure}

Computation of the distance is more complex. First of all, we normalise the length of the whole citation to 1. Next, we align fully matching tokens of two strings to have the same position --- we set it to be an average of their previous positions. Such tokens are called \emph{boundaries}. Beginning and end of a string form additional boundaries. Note that the distance between tokens from a pair forming a boundary is equal to 0. Position of all other tokens is defined relatively to its closest boundaries. An example is provided in Fig. \ref{fig:distance_example}.

\subsubsection{Source matching}
\label{sec:source_matching}

Journal names are abbreviated in the most random way. Usually only a prefix of a word is used (‘appl.’ instead of ‘applied’) or some letters are omitted (‘journal’ becomes ‘jrnl’).That is why we have decided to base source similarity on their LCS (longest common subsequence). We compute character-based LCS of two strings, divide by the length of shorter one and treat the result as similarity.

\subsubsection{Title matching}
\label{sec:title_matching}

The title of an article is not always present in a citation string. When it is, however, we usually need to deal only with spelling errors. That is why a trigram similarity is used in this case.

\subsubsection{Year matching}
\label{sec:year_matching}

All numbers tagged as year are extracted from a citation, their numeric value is computed and for each citation the number closest to 2000 is chosen. The final similarity is a binary value indicating if that numbers are equal.

\subsubsection{Pages matching}
\label{sec:pages_matching}

We extract all numbers tagged as pages, create a set of them for each citation and then compute the ratio of their intersection and sum.

\subsubsection{Whole-string matching}
\label{sec:whole_string_matching}

Additionally three features based on trigram similarity of the whole citation string are used. One of them compares unmodified strings and the other two transformed ones: the first containing only letters and the second only digits from the original strings.

\subsection{Evaluation}
\label{sec:evaluation}

The evaluation was conducted on the CORA-ref data set \cite{McCallumNRS:2000}. It contains several citation clusters (each cluster consisting of citations referencing the same article). We have randomly distributed clusters into 3 slices (see Tab. \ref{tab:slices}.) and used them to perform cross validation. The half of training set for each fold was used in the CRF parser training and the other half was used for the SVM training.

\begin{table}
\centering
\caption{\textbf{Cluster slices.} Some slice statistics are presented along with their usage in particular cross validation phases.}
\label{tab:slices}
\begin{tabular}{l @{\hspace{20pt}} r @{\hspace{5pt}} r @{\hspace{5pt}} r}
\hline\noalign{\smallskip}
 & slice0 & slice1 & slice2 \\
\hline\noalign{\smallskip}
Cluster No. & 80 & 89 & 79 \\
Citation No. & 505 & 596 & 774 \\
Avg. cluster size & 6.31 & 6.70 & 9.80 \\
Max cluster size & 33 & 115 & 121 \\
\hline\noalign{\smallskip}
Parser training & fold0 & fold1 & fold2 \\
Matcher training & fold2 & fold0 & fold1 \\
Matcher testing & fold1 & fold2 & fold0 \\
\noalign{\smallskip}
\hline
\end{tabular}
\end{table}

Then, we have computed pairwise similarities between citation strings. The results were binarised by setting the similarity threshold at 50\%. In order to obtain clusters, a single-link algorithm was applied.

We have performed tests for both versions of author similarity measure (see section \ref{sec:author_matching}.). All the results are presented in Tab. \ref{tab:results_complex} and \ref{tab:results_simple}. The following metrics have been used: 
\begin{itemize}
\item \textbf{cluster recall} -- the percentage of correct clusters that were recovered by a matcher
\item \textbf{pairwise precision} -- the percentage of links returned by a matcher that are correct
\item \textbf{pairwise recall} -- the percentage of correct links that were returned by a matcher
\item \textbf{pairwise $F_1$} -- the harmonic mean of precision and recall
\end{itemize}
As we see, the results are close to those reported in \cite{Poon2007a}. We can also notice that the choice of author similarity  measure only slightly impacts them.

\begin{table}
\centering
\caption{\textbf{Matching results with complex author similarity measure}.}
\label{tab:results_complex}
\begin{tabular}{l @{\hspace{20pt}} r @{\hspace{5pt}} r @{\hspace{5pt}} r @{\hspace{10pt}} r}
\hline\noalign{\smallskip}
 & fold0 & fold1 & fold2 & avg. \\
\hline\noalign{\smallskip}
cluster recall & 65.82\% & 72.50\% & 77.53\% & 71.95\% \\
\noalign{\smallskip}
pairwise precision & 95.21\% & 97.51\% & 94.98\% & 95.90\% \\
pairwise recall & 93.91\% & 93.06\% & 97.43\% & 94.80\% \\
pairwise $F_1$ & 94.56\% & 95.23\% & 96.19\% & 95.33\% \\
\noalign{\smallskip}
\hline
\end{tabular}
\end{table}

\begin{table}
\centering
\caption{\textbf{Matching results with simple author similarity measure}.}
\label{tab:results_simple}
\begin{tabular}{l @{\hspace{20pt}} r @{\hspace{5pt}} r @{\hspace{5pt}} r @{\hspace{10pt}} r}
\hline\noalign{\smallskip}
 & fold0 & fold1 & fold2 & avg. \\
\hline\noalign{\smallskip}
cluster recall & 67.09\% & 76.25\% & 77.53\% & 73.62\% \\
\noalign{\smallskip}
pairwise precision & 94.81\% & 97.45\% & 94.66\% & 95.64\% \\
pairwise recall & 93.03\% & 92.76\% & 97.60\% & 94.46\% \\
pairwise $F_1$ & 93.91\% & 95.05\% & 96.11\% & 95.02\% \\
\noalign{\smallskip}
\hline
\end{tabular}
\end{table}

\section{Heuristic: author indexing}
\label{sec:heuristic_author_indexing}

The main scalability issue in the presented solution is its quadratic runtime. We, therefore, would like to limit the number of necessary pairwise comparisons. The standard approach used in an entity disambiguation is called blocking. The whole set of objects is divided into blocks so that entities are compared pairwise and possibly merged only within a block (cf. \cite{Dendek2013,DendekBL:2012,BolikowskiD:2011}). We have used different method, though.

Bear in mind that the main focus of our system is a very specific type of entity disambiguation: linking citation strings to article metadata stored in the database. That means each citation will be matched to at most one metadata record and records in the store will not be merged. Having observed the above, we have decided to use heuristic based on indexing.

Using an index, metadata records that have the biggest number of author tokens in common  with the examined citation string are retrieved. Author tokens are those describing author name or surname. They were chosen to  be used in our heuristic because we have noticed they are the most reliably parsed part of a citation string. In the following of this section the index is described in more detail.

\subsection{Non-exact matches}
\label{sec:non_exact_matches}

Spelling errors occur commonly in citation strings. That is why an index supporting non-exact matches was desired. We have decided to implement ideas presented by Manning et al.\ in Chapter 3 of \cite{Manning2008} to support retrieval of tokens with edit distance lesser or equal 1. Let us now present this method.

Instead of putting as a key an exact word $w$, we put all the rotations of $w\$$ (where \$ is a character not present in an alphabet). For example, instead of key ‘cat’, keys ‘cat\$’, ‘at\$c’, ‘t\$ca’ and ‘\$cat’ are created. Then, to retrieve a word from the index, we also create all the rotations in a similar manner and for each rotation $r$ of length $n$, all the keys of length $\le n$ that match at least $n-1$ first letters of $r$ and keys of length $\le n+1$ that match first $n$ letters are returned.

For instance, to lookup word ‘cut’, we would create a set of rotations ‘cut\$’, ‘ut\$c’, ‘t\$cu’ and ‘\$cut’. The first three letters of ‘t\$cu’ match ‘t\$ca’, so this key would be retrieved. In the similar manner, to lookup ‘at’, we would have a rotation ‘at\$’ which would match ‘at\$c’.

The whole process can be formalised in the following steps:
\begin{enumerate}
\item Generate all rotations for a token.
\item For each rotation $r$ find all matching rotations in the index:
\begin{enumerate}
\item Let $r=bc$, where $|c|=1$
\item Find in the index the lexicographically smallest token $t$, such that $b$ is prefix of $t$
\item Scan the following index entries to retrieve all words beginning with $b$ of max length $|r|$ and all beginning with $r$ of max length $|r+1|$ with their document IDs
\end{enumerate}
\item Flatten document ID lists and convert them to a set.
\end{enumerate}

\subsection{Heuristic summary}
\label{sec:heuristic_summary}

To conclude, let us present the whole heuristic matching process:
\begin{enumerate}
\item Extract all author tokens from the citation
\item For each token, retrieve all matching documents
\item For each document retrieved, compute the number of matching tokens
\item Filter out documents containing less than $\max(1, M-1)$ matching authors, where $M=$ the maximum number of matching authors
\item All remaining documents heuristicly match the citation
\end{enumerate}

\section{Hadoopisation}
\label{sec:hadoopisation}

Apache Hadoop is the most notable implementation of MapReduce paradigm. In this section we present how to implement the algorithms described above using MapReduce and Hadoop environment. To do that accurately, though, we need to describe some technicalities first. Even more technical details not covered in this article are discussed in \cite{DendekCFKWB:2013}.

Hadoop \emph{SequenceFile} is a binary file containing a list of key-value pairs. A SequenceFile with sorted keys enriched by additional index file enabling fast record retrieval is called a \emph{MapFile}. This data structure is used to store our index. Additionally, we assume our input and output data will be stored as SequenceFiles.

\subsection{Index building}
\label{sec:index_building}

\begin{figure}
\begin{center}
\includegraphics[scale=0.2]{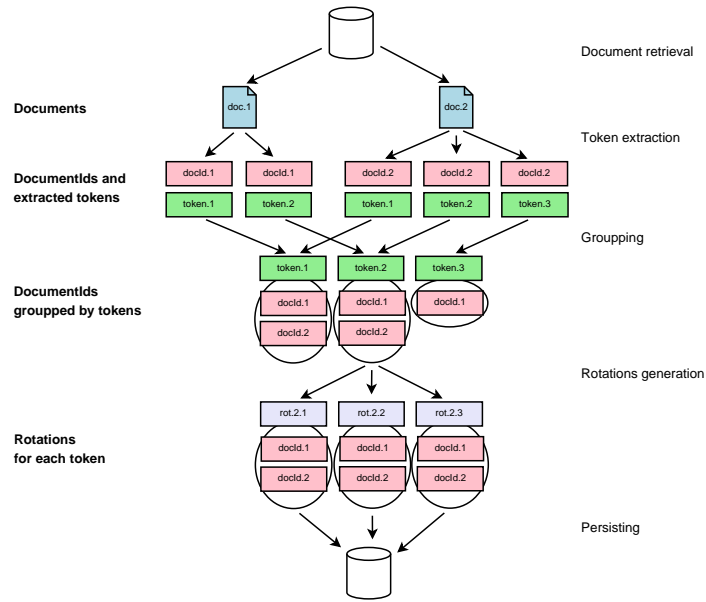}
\end{center} 
\caption{\textbf{Index building.} The documents are read from a SequenceFile and all author tokens (tokenised names) and document IDs are extracted from the metadata. Then, document IDs with the same tokens are grouped. Eventually, for each token, its rotations are generated and everything is persisted in a SequenceFile.}
\label{fig:cit-match-index-building-vertical}
\end{figure}

Let us begin with the presentation of the index building process. The most crucial part is presented in Fig. \ref{fig:cit-match-index-building-vertical}. Note that steps transforming one entry into many can be implemented as map tasks (i.e. token extraction and rotations generation) and those transforming many into one as reduce (i.e. grouping). 

The process depicted generates all necessary entries and stores them as a SequenceFile. To sort it and transform into a MapFile Hadoop built-in functions are used.

\subsection{Actual matching}
\label{sec:actual_matching}

\begin{figure}
\begin{center}
\includegraphics[width=\linewidth]{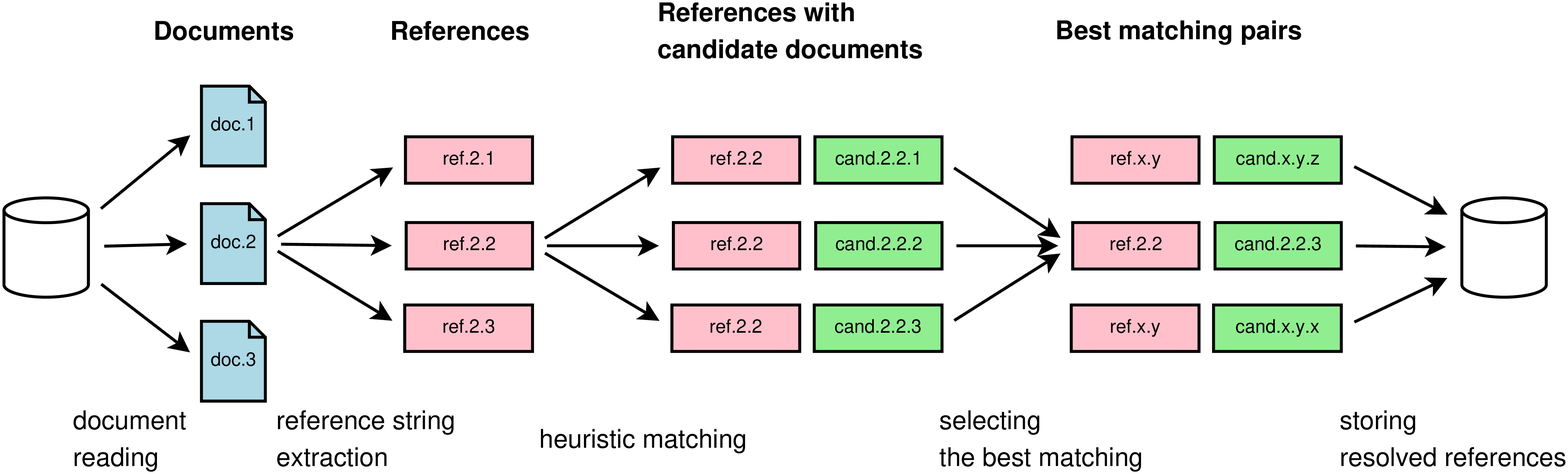}
\end{center} 
\caption{\textbf{Citation matching steps.} The documents are read from an appropriate SequenceFile and references are extracted from their metadata. Then, the actual matching occurs: in the first step heuristic is used to find documents that may match citations, in the next the best match is selected for each citation. Eventually, the results are persisted in a SequenceFile.}
\label{fig:cit-match-main}
\end{figure}

Having built the index, we can step to the actual matching phase. It is presented in Fig \ref{fig:cit-match-main}. Here, the reference extraction and the heuristic matching are done in map steps. Choosing the best matching document is achieved by selecting the one with the biggest similarity to the citation string, what is done as a reduce step. The similarity is computed using metrics defined in section \ref{sec:metadata_fields_matching} with whole-string features omitted and the simple version of author matching.

\subsection{Speed evaluation}
\label{sec:speed_evaluation}

We have evaluated efficiency of our solution using PMC Open Access Subset document set. It consists of over 450~thousand documents containing 12~million citations. 

The benchmark was performed on our Hadoop cluster \cite{Kawa2013} which consists of a four "fat" slave nodes and a virtual machine on a separate physical machine in the role of NameNode, JobTracker and HBase master. Each worker node has four AMD Opteron 6174 processors (48 cores in total), 192 GB of RAM, four 600 GB disks which work in RAID 5 array. 

The time spent in each phase is presented in the Table \ref{tab:speed_results}.

\begin{table}
\centering
\caption{\textbf{The time spent in individual phases}.}
\label{tab:speed_results}
\begin{tabular}{ |l|l|c|c|c| }
\hline
\multicolumn{2}{ |c| }{\multirow{2}{*}{Phase}} & \multirow{2}{*}{Time spent} & \multicolumn{2}{ |c| }{Task No.} \\
\cline{4-5}
\multicolumn{2}{ |c| }{} & & Map & Reduce \\
\hline
\multirow{1}{*}{Index building} & All & 0:00:57 & 13 & 2 \\
\hline
\multirow{3}{*}{Matching} & Citation extraction & 0:00:46 & 13 & 0 \\
  & Heuristic matching & 3:01:38 & 745 & 0 \\
  & Selecting the best match & 2:51:00 & 996 & 1 \\ 
\hline
\end{tabular}
\end{table}

\section{Conclusions and future work}
\label{sec:conclusions}

In this paper we have presented an efficient citation matching solution using Apache Hadoop. After developing a basic citation matching technique, we have shown how to scale it up to handle millions of citations. In particular, we have presented a way of creating an approximate index using the MapReduce paradigm.

The big data enables new, unparalleled possibilities, which need more research. It is especially worth investigating how our model training phase may benefit from huge amounts of data. On the other hand, when dealing with enormous training set, perhaps many of examples are not relevant. Maybe we want to select only the most important ones. If so, how do we do that? Only the further investigation may answer these questions.

\bibliographystyle{splncs}
\bibliography{bibliography}
 
\end{document}